\documentclass[preprintnumbers]{revtex4}
\usepackage{amssymb}
\usepackage{amsmath}
\usepackage{graphicx}
\usepackage{calligra}
\usepackage{mathrsfs}
\usepackage{dcolumn}
\usepackage{bm}
\usepackage{subfigure}
\usepackage{color}
\usepackage{CJKutf8}
\numberwithin{equation}{section}
\numberwithin{figure}{section}

\begin{document}
\title{Universality on thermodynamic relation with corrections in Einstein-Bel-Robinson gravity Black hole}
\author{Hai-Long Zhen,$^{1,2}$ Huai-Fan Li,$^{2,3}$ Yu-Bo Ma$^{1,2}$}
\thanks{\emph{e-mail: yuboma.phy@gmail.com}(corresponding author)}

\address{$^1$Department of Physics, Shanxi Datong University, Datong 037009, China\\
$^2$Institute of Theoretical Physics, Shanxi Datong University, Datong, 037009, China\\
$^3$Shanxi College of Technology, College of General Education, Shuozhou 036000, China\\}

\begin{abstract}

The generalized thermodynamic extremum relation, as proposed by Goon and Penco, establishes a novel theoretical framework for the study of spacetime thermodynamics. However, extant investigations generally assume that the black hole state parameter is solely a first-order function of the perturbation parameter when exploring the Goon-Penco relation in diverse spacetime contexts. An analytic expression for the perturbation parameter as a function of the black hole entropy can be expressed by treating the black hole mass as constant. The present study addresses this limitation and provides insight into the universal Goon-Penco relation when multiple thermodynamic state parameters behave as higher order functions of the perturbation parameters. Notably, we have not only established a universal relational formula in the case of multiple state variables, but more importantly, we have put forward an innovative conjecture that reveals the existence of a universal relation between displaced thermodynamic quantities in spacetime in the context of an arbitrary black hole. These theoretical breakthroughs are expected to open up new exploration directions for quantum gravity research.
\end{abstract}

\maketitle

\section{Introduction}\label{one}

The Weak Gravity Conjecture (WGC) establishes a fundamental constraint on the theoretical framework of quantum gravity. This conjecture posits that the canonical interactions of specific states with some lighter charged particle or field must exceed the gravitational interactions within any self-consistent quantum gravity theory \cite{C0509212,N0601001}. Specifically, it asserts the existence of at least one particle state with charge $Q$ and mass $M$ that satisfies the inequality relation $Q$$>$$M$, thereby ensuring that gravitational attraction is weaker than the electromagnetic repulsion. However, for extremal black holes(BH) saturating the bound for a charged BH, the charge-to-mass ratio $Q=M$. This theoretical conjecture is of particular significance in physical systems with global symmetry absenting \cite{T084019}, and its theoretical foundations are rooted in the principles of quantum gravity \cite{D08380}. Utilizing the research framework of the Weak Gravity Conjecture (WGC), Goon and Penco have pioneered the study of the universal thermodynamic correlation between black hole entropy and extremality under perturbation parameter \cite{G101103}. The study that perturbation corrections to the free energy induce novel relationship between mass, temperature, and entropy is demonstrated. Through a higher-order expansion of the perturbation parameters, a theoretically significant approximation that can be naturally connected to the theory of higher-order derivative corrections is identified \cite{Y00772}. A rigorous theoretical connection between shifts in entropy and the charge-to-mass ratio is yielded \cite{C046003,Y0606100,C08546}. Based on the relation proposed by Goon-Penco, a series of significant advancements have been made. The thermodynamic relations have systematically been analyzed in various AdS spacetimes, including but not limited to: charged BTZ black holes, high-dimensional Reissner-Nordstr$\ddot{o}$m (A)dS black holes \cite{P14324,J139149}, Kerr-Newman (A)dS black holes \cite{Y035106}, Power-law corrected AdS black holes \cite{J17014}, as well as the Charged-Rotating-AdS black holes surrounded by Quintessence and String Clouds \cite{A07079} from the perspective of the WGC \cite{J116581,S11161,P08530,M13784,J025305,C116686,L08527,S115279,H11535,M09654,K02639,L00015,BL081901,J169168}. These studies not only verify the universality of the Goon-Penco relation, but also provide new theoretical perspectives on understanding the deep connection between quantum gravity and black hole thermodynamics.

However, extant studies have the following important limitations: first, all the models that have been explored are restricted to the spacetime gauge containing only first-order term of the perturbation parameter $\eta$; second, when calculating the Goon-Penco (GP) relation for the complex black holes of multiple parameters, it is common to consider only a certain pair of thermodynamic state parameters as a function of the perturbation parameter $\eta$, while the other thermodynamic quantities are approximated as constants. In the more general case, where the gauge of spacetime contains higher-order terms for the perturbation parameter $\eta$ and multiple thermodynamic state parameters are functions of $\eta$ simultaneously, the challenge of obtaining an analytic expression for the perturbation parameter $\eta$ makes the study of the GP relation in this type of spacetime context a major theoretical challenge. In this study, we methodically explore the establishment of the GP relation using the black hole in Einstein-Bel-Robinson gravity theory as a specific model. The model exhibits two distinctive characteristics: Firstly, the black hole thermodynamic quantities contain higher-order terms of the perturbation parameter, $\eta$. Secondly, all the thermodynamic state variables are functions of the perturbation, $\eta$. These characteristics make our study universal. Moreover, the proposed research method effectively overcomes the conventional limitations for studying GP relations within a complex spacetime context. This approach is not only applicable to specific models, but also has a wide generalization value, providing a new research paradigm for a deeper understanding of thermodynamic polar relations in strong gravitational fields.

The paper is arranged as follows: In Sec. \ref{two}, the expression for the thermodynamic state parameters of black holes in the Einstein-Bel-Robinson theory of gravity is presented in order to maintain the coherence of the theoretical discussion. In particular, the key thermodynamic variables such as black hole energy $M$, entropy $S$, and temperature $T$ as the higher-order functions of the perturbation parameter $\eta$, are demonstrated in detail. In Sec. \ref{three}, the GP relation for Einstein-Bel-Robinson black holes is systematically investigated using three diverse methods: (1) the direct calculation method: The GP relation is obtained by calculating both the black hole energy $M$ and entropy $S$ as higher-order functions of the perturbation parameter $\eta$. (2) composite function derivation method: the GP relation is discussed by composite function derivation method. (3) full differential method: the universal GP relation is deduced when the black hole energy $M$ is a function of multiple thermodynamic state parameters, and each state parameter is a higher-order function of the perturbation parameter $\eta$, under the condition that the thermodynamic quantities at the black hole horizon satisfy the first law of thermodynamics. The universality of the relation proposed by Goon and Penco is independent of the specific functional form of each thermodynamic quantity as a function of the perturbation parameter $\eta$. Finally, A brief summary is presented in Sec. \ref{four}

\section{4D Einstein-Bel-Robinson Gravity} \label{two}

In $4$ dimensions, EBR gravity is determined by the following action \cite{S13172,G08254}
\begin{align}\label{2.1}
I&=\frac{1}{16\pi G}\int{{{d}^{4}}}x\sqrt{g}\left[ R-2\Lambda -\beta \left( P_{4}^{2}-E_{4}^{2} \right) \right].
\end{align}
Where $R$ is the Ricci scalar, $\beta$ is the coupling constant of the theory, and $\Lambda=-3/l^2$, where $l$ refers to the curvature radius of the maximally symmetric AdS solution of the field equations. Where $P_{4}^{2}$ and $E_{4}^{2}$ are the Pontryagin and the Euler topological densities which are related to the Bel-Robinson tensor ${{T}_{\mu \nu \lambda \rho }}$ in four dimensions by means of the relation [31,32]
\begin{align}\label{2.2}
{{T}^{\mu \nu \lambda \rho }}{{T}_{\mu \nu \lambda \rho }}&=\frac{1}{4}(P_{4}^{2}-E_{4}^{2}).
\end{align}
The Bel-Robinson tensor is defined as:
\begin{align}\label{2.3}
{{T}^{\mu \nu \delta \sigma }}&={{R}^{\mu \rho \gamma \delta }}R_{\rho \gamma }^{\nu \sigma }+{{R}^{\mu \rho \gamma \sigma }}R_{\rho \gamma }^{\nu \delta }-\frac{1}{2}{{g}^{\mu \nu }}{{R}^{\rho \gamma \alpha \delta }}R_{\rho \gamma \alpha }^{\sigma }.
\end{align}
Varying the action (\ref{2.1}) with respect to the metric yields the following equations of motion
\begin{align}\label{2.4}
{{\varepsilon }_{ab}}&={{R}_{ab}}-\frac{1}{2}{{g}_{ab}}R+\Lambda {{g}_{ab}}-\beta {{\kappa }_{ab}}=0,
\end{align}
where
\begin{align}\label{2.5}
{{\kappa }_{ab}}&=\frac{1}{2}{{g}_{ab}}E_{4}^{2}-2[2{{E}_{4}}R{{R}_{ab}}-4{{E}_{4}}R_{a}^{\rho }{{R}_{b\rho }}+2{{E}_{4}}R_{a}^{\rho \sigma \lambda }{{R}_{b\rho \sigma \lambda }}+4{{E}_{4}}{{R}_{\rho \sigma }}{{R}_{a\rho \sigma b}}+2{{g}_{ab}}RW {{E}_{4}} \notag \\
&-2R{{\nabla }_{a}}{{\nabla }_{b}}{{E}_{4}}-4{{R}_{ab}}W {{E}_{4}}+4({{R}_{ab}}{{\nabla }^{\rho }}{{\nabla }_{b}}{{E}_{4}}+{{R}_{b\rho }}{{\nabla }^{\rho }}{{\nabla }_{b}}{{E}_{4}})-4{{g}_{ab}}{{R}_{\rho \sigma }}{{\nabla }^{\sigma }}{{\nabla }^{\rho }}{{E}_{4}}+4{{R}_{a\rho b\sigma }}{{\nabla }^{\sigma }}{{\nabla }^{\rho }}{{E}_{4}}].
\end{align}
We added a small perturbation parameter $\eta$ to the cosmological constant in the action \cite{G101103,P14324,Y035106}. By taking limit of $\eta$ to zero, we can reproduce the original form of the action, modifying the action to
\begin{align}\label{2.6}
I&=\frac{1}{16\pi G}\int{{{d}^{4}}}x\sqrt{g}\left[ R-2(1+\eta )\Lambda -\beta \left( P_{4}^{2}-E_{4}^{2} \right) \right].
\end{align}
Thus, the shift of action can be determined as
\begin{align}\label{2.7}
\vartriangle I&=-\frac{1}{16\pi G}\int{{{d}^{4}}}x\sqrt{g}(2\eta \Lambda ).
\end{align}
The most general radially symmetric metric can be written in the form \cite{R137690,S01078}
\begin{align}\label{2.8}
d{{s}^{2}}&=-N(r)f(r)d{{t}^{2}}+\frac{d{{r}^{2}}}{f(r)}+{{r}^{2}}\left( d{{\theta }^{2}}+\frac{{{\sin }^{2}}(\sqrt{k\theta })}{k}d{{\phi }^{2}} \right),
\end{align}
where the metric functions $N(r)$ and $f(r)$ must be determined from the field Eq. (\ref{2.4}) and $k\in \{1,0,-1\}$ respectively corresponding to spherical, planar, and hyperbolic transverse sections. Note that in general $N(r)$ is not constant, which is the generic situation for higher curvature gravity theories. The exception to this is the class of generalized quasitopological gravity theories(GQGTs) \cite{P015004,R064055,P104005}, whose field equations ensure that $N(r)$ is a constant that can be set to unity without loss of generality.

Solving the $tt$ and $rr$ components of Eq. (\ref{2.4}), yields [32]
\begin{align}\label{2.9}
N(r)&=1-\left( \frac{896(1+\eta )\Lambda {{m}^{2}}}{{{r}^{6}}}+\frac{3584m3}{{{r}^{9}}} \right)\beta \notag \\
&+\left( \frac{372736{{(1+\eta )}^{4}}{{\Lambda }^{4}}{{m}^{2}}}{9{{r}^{6}}} \right.-\frac{3686k{{(1+\eta )}^{3}}{{\Lambda }^{3}}{{m}^{2}}}{{{r}^{8}}}+\frac{10452992{{(1+\eta )}^{3}}{{\Lambda }^{3}}{{m}^{3}}}{9{{r}^{9}}} \notag \\
&-\frac{17694720k{{(1+\eta )}^{2}}{{\Lambda }^{2}}{{m}^{3}}}{11{{r}^{11}}}-\frac{8019968{{(1+\eta )}^{2}}{{\Lambda }^{2}}{{m}^{4}}}{12}+\frac{50429952k(1+\eta )\Lambda {{m}^{4}}}{{{r}^{14}}} \notag \\
&-\frac{86792016k(1+\eta )\Lambda {{m}^{5}}}{5{{r}^{15}}}+\frac{4236115968k{{m}^{5}}}{17{{r}^{17}}}\left. -\frac{569393152{{m}^{6}}}{{{r}^{18}}} \right){{\beta }^{2}}.
\end{align}
\begin{align}\label{2.10}
f(r)&=k-\frac{(1+\eta )\Lambda {{r}^{2}}}{3}-\frac{2m}{r}+ \notag \\
&+\left( -\frac{32{{(1+\eta )}^{4}}{{\Lambda }^{4}}{{r}^{2}}}{27} \right.-\frac{896{{(1+\eta )}^{2}}{{\Lambda }^{2}}{{m}^{2}}}{3{{r}^{4}}}+\frac{768(1+\eta )\Lambda k{{m}^{2}}}{{{r}^{6}}}-\frac{3200(1+\eta )\Lambda {{m}^{2}}}{{{r}^{7}}} \notag \\
&\left. +\frac{4608k{{m}^{3}}}{{{r}^{9}}}-\frac{8576{{m}^{4}}}{{{r}^{10}}} \right)\beta +\left( -\frac{4096{{(1+\eta )}^{7}}{{\Lambda }^{7}}{{r}^{2}}}{243} \right. \notag \\
&+\frac{114688{{(1+\eta )}^{5}}{{\Lambda }^{5}}{{m}^{2}}}{9{{r}^{4}}}-\frac{475136{{(1+\eta )}^{4}}{{\Lambda }^{4}}{{k}^{2}}}{3{{r}^{6}}}+\frac{4280320{{(1+\eta )}^{4}}{{\Lambda }^{4}}{{m}^{3}}}{9{{r}^{7}}}\notag \\
&+\frac{327680{{(1+\eta )}^{3}}{{\Lambda }^{3}}{{k}^{2}}{{m}^{2}}}{{{r}^{8}}}-\frac{2465792{{(1+\eta )}^{3}}{{\Lambda }^{3}}k{{m}^{3}}}{{{r}^{9}}}-\frac{3162112{{(1+\eta )}^{3}}{{\Lambda }^{3}}{{m}^{4}}}{9{{r}^{10}}}\notag \\
&+\frac{1966080{{(1+\eta )}^{2}}{{\Lambda }^{2}}{{k}^{2}}{{m}^{3}}}{{{r}^{11}}}+\frac{219955200{{(1+\eta )}^{2}}{{\Lambda }^{2}}{{k}^{4}}}{11{{r}^{12}}}-\frac{236548096{{(1+\eta )}^{2}}{{\Lambda }^{2}}{{m}^{5}}}{3{{r}^{13}}}\notag \\
&-\frac{47185920(1+\eta )\Lambda {{k}^{2}}{{m}^{4}}}{{{r}^{14}}}+\frac{3749331456(1+\eta )\Lambda {{k}^{2}}{{m}^{5}}}{{{r}^{15}}}-\frac{29367336(1+\eta )\Lambda {{m}^{6}}}{5{{r}^{16}}}\notag \\
&-\frac{47185920(1+\eta )\Lambda {{k}^{2}}{{m}^{4}}}{{{r}^{14}}}+\frac{3749331456(1+\eta )\Lambda {{k}^{2}}{{m}^{5}}}{{{r}^{15}}}-\frac{29367336(1+\eta )\Lambda {{m}^{6}}}{5{{r}^{16}}}\notag \\
&\left. -\frac{283115520{{k}^{2}}{{m}^{5}}}{{{r}^{17}}}+\frac{20514373632k{{m}^{6}}}{17{{r}^{18}}}-\frac{1275707392{{m}^{7}}}{{{r}^{19}}} \right){{\beta }^{2}},
\end{align}
where $m$ is a constant of integration. As we introduced a perturbation to the action, the thermodynamic variables also shift to their perturbed forms at the theoretical level. These variables can then be calculated using the following metric:
\begin{align}\label{2.11}
T&=\frac{1}{4\pi {{r}_{+}}}-\frac{(1+\eta )\Lambda }{4\pi }\notag \\
&+\frac{\beta }{\pi }\left( -\frac{12(1+\eta )\Lambda }{r_{+}^{5}}+\frac{8{{(1+\eta )}^{2}}{{\Lambda }^{2}}}{r_{+}^{3}}-\frac{28{{(1+\eta )}^{3}}{{\Lambda }^{3}}}{27{{r}_{+}}}-\frac{2}{r_{+}^{7}}-\frac{26{{(1+\eta )}^{4}}{{\Lambda }^{4}}}{27{{r}_{+}}} \right)+o({{\beta }^{2}}).
\end{align}
\begin{align}\label{2.12}
M&=\frac{{{r}_{+}}}{2}-\frac{(1+\eta )\Lambda r_{+}^{3}}{6}\notag \\
&+\beta \left( \frac{20}{r_{+}^{5}}-\frac{104(1+\eta )\Lambda }{3r_{+}^{3}}+\frac{16{{(1+\eta )}^{2}}{{\Lambda }^{2}}}{{{r}_{+}}}-\frac{56{{(1+\eta )}^{3}}{{\Lambda }^{3}}{{r}_{+}}}{27}-\frac{52{{(1+\eta )}^{4}}{{\Lambda }^{4}}r_{+}^{3}}{81} \right)+o({{\beta }^{2}}).
\end{align}
\begin{align}\label{2.13}
S&=\pi r_{+}^{2}+\frac{32\pi \beta }{r_{+}^{4}}(3-2(1+\eta )\Lambda r_{+}^{2}+{{(1+\eta )}^{2}}{{\Lambda }^{2}}r_{+}^{4})+o({{\beta }^{2}}).
\end{align}
In the context of extended phase space thermodynamics, or what is more commonly known as black hole chemistry \cite{D2126,D0559,D063001,R6233} the cosmological constant can be interpreted as thermodynamic pressure via
\begin{align}\label{2.14}
P&=-\frac{\Lambda }{8\pi },
\end{align}
with conjugate volume
\begin{align}\label{2.15}
{{V}_{\Lambda }}&=\frac{4\pi r_{+}^{3}}{3}+\beta \pi \left( \frac{448}{3r_{+}^{3}}-\frac{704{{(1+\eta )}^{2}}{{\Lambda }^{2}}{{r}_{+}}}{9}+\frac{1664{{(1+\eta )}^{3}}{{\Lambda }^{3}}r_{+}^{3}}{81} \right)+o({{\beta }^{2}}).
\end{align}
From Eqs. (\ref{2.11}), (\ref{2.12}) and (\ref{2.13}), the energy, $M$, and entropy, $S$, of the black hole are functions of ${{r}_{+}}$, $P$, $\beta$ and perturbation parameter $\eta$, i.e.
$M=M({{r}_{+}},P,\beta ,\eta )$, $S=S({{r}_{+}},P,\beta ,\eta )$

\section{Goon-Penco relation} \label{three}

The GP relation was constructed using mass, temperature, entropy, and the perturbation parameter $\eta$ \cite{G101103,P14324,Y035106}. in addition, each side of the relation includes a partial derivative term of mass and entropy with respect to $\eta$,
\begin{align}\label{3.1}
\frac{\partial {{M}_{ext}}(\mathcal{Q},\eta )}{\partial \eta }&=\underset{M\to {{M}_{ext}}}{\mathop{\lim }}\,-T{{\left( \frac{\partial S}{\partial \eta } \right)}_{M,\mathcal{Q}}},
\end{align}
where $M$ and $\mathcal{Q}$ represent the mass and additional quantities.

$1)$ direct method of calculation

Under the condition that $P$ and $\beta$ are constant, when $M\to {{M}_{0}}$, the position of the black hole horizon, $r_+$, is a function of perturbation parameter, $\eta$, we can obtain
\begin{align}\label{3.2}
0&=-\left[ \frac{\Lambda r_{+}^{3}}{6}-\beta \left( -\frac{104\Lambda }{3r_{+}^{3}}+\frac{32(1+\eta ){{\Lambda }^{2}}}{{{r}_{+}}}-\frac{56{{(1+\eta )}^{2}}{{\Lambda }^{3}}{{r}_{+}}}{279}-\frac{208{{(1+\eta )}^{3}}{{\Lambda }^{4}}r_{+}^{3}}{81} \right) \right] \notag \\
&+\left[ \frac{1}{2}-\frac{(1+\eta )\Lambda r_{+}^{2}}{2} \right.
+\beta \left. \left( -\frac{100}{r_{+}^{6}}+\frac{104(1+\eta )\Lambda }{r_{+}^{4}}-\frac{16{{(1+\eta )}^{2}}{{\Lambda }^{2}}}{r_{+}^{2}}-\frac{56{{(1+\eta )}^{3}}{{\Lambda }^{3}}}{27}-\frac{52{{(1+\eta )}^{4}}{{\Lambda }^{4}}r_{+}^{2}}{27} \right) \right]\frac{\partial {{r}_{+}}}{\partial \eta }.
\end{align}
It is evident that, under the condition that $P$, $\beta$ and $S$ are constant, subject to Eq. (\ref{2.13}), the position of the black hole horizon, $r_+$, varies with the perturbation parameter ,$\eta$. That is to say, $r_+$ is a function of $\eta$.
\begin{align}\label{3.3}
0&=\frac{32\pi \beta }{r_{+}^{4}}(-2\Lambda r_{+}^{2}+2(1+\eta ){{\Lambda }^{2}}r_{+}^{4})+\left( 2\pi {{r}_{+}}+128\pi \beta \left( -\frac{3}{r_{+}^{5}}+\frac{(1+\eta )\Lambda }{r_{+}^{3}} \right) \right)\frac{\partial {{r}_{+}}}{\partial \eta }.
\end{align}
According to the derivation law of composite function, we obtain the following result:
\begin{align}\label{3.4}
{{\left( \frac{\partial M}{\partial \eta } \right)}_{P,\beta ,S}}&={{\left( \frac{\partial M}{\partial \eta } \right)}_{P,\beta ,{{r}_{+}}}}+{{\left( \frac{\partial M}{\partial {{r}_{+}}} \right)}_{P,\beta ,\eta }}{{\left( \frac{\partial {{r}_{+}}}{\partial \eta } \right)}_{P,\beta ,S}}.
\end{align}
\begin{align}\label{3.5}
\underset{M\to {{M}_{0}}}{\mathop{\lim }}\,{{\left( \frac{\partial S}{\partial \eta } \right)}_{P,\beta ,M}}&=\underset{M\to {{M}_{0}}}{\mathop{\lim }}\,\left[ {{\left( \frac{\partial S}{\partial \eta } \right)}_{P,\beta ,{{r}_{+}}}}+{{\left( \frac{\partial S}{\partial {{r}_{+}}} \right)}_{P,\beta ,\eta }}{{\left( \frac{\partial {{r}_{+}}}{\partial \eta } \right)}_{P,\beta ,M}} \right].
\end{align}
Substituting Eqs. (\ref{3.2}) and (\ref{3.3}) into Eqs. (\ref{3.4}) and (\ref{3.5}), respectively, yields
\begin{align}\label{3.6}
{{\left( \frac{\partial {{M}_{0}}}{\partial \eta } \right)}_{P,\beta ,S}}&=-\left[ \frac{\Lambda r_{+}^{3}}{6}-\beta \left( -\frac{104\Lambda }{3r_{+}^{3}}+\frac{32(1+\eta ){{\Lambda }^{2}}}{{{r}_{+}}}-\frac{56{{(1+\eta )}^{2}}{{\Lambda }^{3}}{{r}_{+}}}{279}-\frac{208{{(1+\eta )}^{3}}{{\Lambda }^{4}}r_{+}^{3}}{81} \right) \right]\notag \\
&-\frac{\frac{32\pi \beta }{r_{+}^{4}}(-2\Lambda r_{+}^{2}+2(1+\eta ){{\Lambda }^{2}}r_{+}^{4})}{2\pi {{r}_{+}}\left( 1+64\beta \left( -\frac{3}{r_{+}^{6}}+\frac{(1+\eta )\Lambda }{r_{+}^{4}} \right) \right)}\left[ \frac{1}{2}-\frac{(1+\eta )\Lambda r_{+}^{2}}{2} \right.\notag \\
&+\beta \left. \left( -\frac{100}{r_{+}^{6}}+\frac{104(1+\eta )\Lambda }{r_{+}^{4}}-\frac{16{{(1+\eta )}^{2}}{{\Lambda }^{2}}}{r_{+}^{2}}-\frac{56{{(1+\eta )}^{3}}{{\Lambda }^{3}}}{27}-\frac{52{{(1+\eta )}^{4}}{{\Lambda }^{4}}r_{+}^{2}}{27} \right) \right]\notag \\
&=-\left[ \frac{\Lambda r_{+}^{3}}{6}-\beta \left( -\frac{104\Lambda }{3r_{+}^{3}}+\frac{32(1+\eta ){{\Lambda }^{2}}}{{{r}_{+}}}-\frac{56{{(1+\eta )}^{2}}{{\Lambda }^{3}}{{r}_{+}}}{279}-\frac{208{{(1+\eta )}^{3}}{{\Lambda }^{4}}r_{+}^{3}}{81} \right) \right]\notag \\
&-A-\frac{\frac{32\pi \beta }{r_{+}^{4}}(-2\Lambda r_{+}^{2}+2(1+\eta ){{\Lambda }^{2}}r_{+}^{4})}{2\pi {{r}_{+}}\left( 1+64\beta \left( -\frac{3}{r_{+}^{6}}+\frac{(1+\eta )\Lambda }{r_{+}^{4}} \right) \right)}B,
\end{align}
\begin{align}\label{3.7}
\underset{M\to {{M}_{0}}}{\mathop{\lim }}\,{{\left( \frac{\partial S}{\partial \eta } \right)}_{P,\beta ,M}}&=\frac{32\pi \beta }{r_{+}^{4}}(-2\Lambda r_{+}^{2}+2(1+\eta ){{\Lambda }^{2}}r_{+}^{4})+2\pi {{r}_{+}}\left( +64\beta \left( -\frac{3}{r_{+}^{6}}+\frac{(1+\eta )\Lambda }{r_{+}^{4}} \right) \right)\frac{A}{B},
\end{align}
with
\begin{align}\label{3.8}
A&=\left[ \frac{\Lambda r_{+}^{3}}{6}-\beta \left( -\frac{104\Lambda }{3r_{+}^{3}}+\frac{32(1+\eta ){{\Lambda }^{2}}}{{{r}_{+}}}-\frac{56{{(1+\eta )}^{2}}{{\Lambda }^{3}}{{r}_{+}}}{279}-\frac{208{{(1+\eta )}^{3}}{{\Lambda }^{4}}r_{+}^{3}}{81} \right) \right],\notag \\
B&=\left[ \frac{1}{2}-\frac{(1+\eta )\Lambda r_{+}^{2}}{2} \right.+\beta \left. \left( -\frac{100}{r_{+}^{6}}+\frac{104(1+\eta )\Lambda }{r_{+}^{4}}-\frac{16{{(1+\eta )}^{2}}{{\Lambda }^{2}}}{r_{+}^{2}}-\frac{56{{(1+\eta )}^{3}}{{\Lambda }^{3}}}{27}-\frac{52{{(1+\eta )}^{4}}{{\Lambda }^{4}}r_{+}^{2}}{27} \right) \right]
\end{align}
From Eqs. (\ref{3.6}) and (\ref{3.7}), we obtain
\begin{align}\label{3.9}
&2\pi {{r}_{+}}\left( 1+64\beta \left( -\frac{3}{r_{+}^{6}}+\frac{(1+\eta )\Lambda }{r_{+}^{4}} \right) \right){{\left( \frac{\partial {{M}_{0}}}{\partial \eta } \right)}_{P,\beta ,S}} \notag \\
&=-2\pi {{r}_{+}}\left( 1+64\beta \left( -\frac{3}{r_{+}^{6}}+\frac{(1+\eta )\Lambda }{r_{+}^{4}} \right) \right)A-\frac{32\pi \beta }{r_{+}^{4}}(-2\Lambda r_{+}^{2}+2(1+\eta ){{\Lambda }^{2}}r_{+}^{4})B\notag \\
\underset{M\to {{M}_{0}}}{\mathop{\lim }}\,B{{\left( \frac{\partial S}{\partial \eta } \right)}_{P,\beta ,M}}&=\frac{32\pi \beta }{r_{+}^{4}}(-2\Lambda r_{+}^{2}+2(1+\eta ){{\Lambda }^{2}}r_{+}^{4})B+2\pi {{r}_{+}}\left( 1+64\beta \left( -\frac{3}{r_{+}^{6}}+\frac{(1+\eta )\Lambda }{r_{+}^{4}} \right) \right)A.
\end{align}
\begin{align}\label{3.10}
2\pi {{r}_{+}}\left( 1+64\beta \left( -\frac{3}{r_{+}^{6}}+\frac{(1+\eta )\Lambda }{r_{+}^{4}} \right) \right){{\left( \frac{\partial {{M}_{0}}}{\partial \eta } \right)}_{P,\beta ,S}}+\underset{M\to {{M}_{0}}}{\mathop{\lim }}\,B{{\left( \frac{\partial S}{\partial \eta } \right)}_{P,\beta ,M}}&=0.
\end{align}
From Eqs. (\ref{3.10}) and (\ref{2.11}), we obtain
\begin{align}\label{3.11}
{{\left( \frac{\partial {{M}_{0}}}{\partial \eta } \right)}_{P,\beta ,S}}&=-\underset{M\to {{M}_{0}}}{\mathop{\lim }}\,\frac{B}{2\pi {{r}_{+}}\left( 1+64\beta \left( -\frac{3}{r_{+}^{6}}+\frac{(1+\eta )\Lambda }{r_{+}^{4}} \right) \right)}{{\left( \frac{\partial S}{\partial \eta } \right)}_{P,\beta ,M}}=-\underset{M\to {{M}_{0}}}{\mathop{\lim }}\,T{{\left( \frac{\partial S}{\partial \eta } \right)}_{P,\beta ,M}},
\end{align}
which is a form of the GP relation (\ref{3.1}). In the aforementioned discussion, it is posited that, within the context of constant energy $M$, the perturbation parameter $\eta$, when varied, induces solely a modification in the horizon position $r_+$ of the black hole, while leaving $P$ unaltered. It can be determined from Eq. (\ref{2.12}) that, under the condition that the energy $M$ is constant, the perturbation parameter $\eta$ can be altered to ensure the validity of the equation. This modification may entail a change in either the horizon position $r_+$ or the perturbation parameter $\eta$, or both $r_+$ and $\eta$ provided that equation (2.12) is satisfied.

$2)$ The Derivation of Composite Function

In the context of the Eq. (\ref{2.12}), it can be demonstrated that, under the condition that $\beta$ and $S$ remain constant, and when $M\rightarrow{M_0}$, both the black hole horizon position $r_+$ and $P$ are functions of $\eta$. According to the law of derivation of the composite function, we obtain
\begin{align}\label{3.12}
{{\left( \frac{\partial M}{\partial \eta } \right)}_{P(\eta ),\beta ,S}}&={{\left( \frac{\partial M}{\partial \eta } \right)}_{P(\eta ),\beta ,{{r}_{+}}}}+{{\left( \frac{\partial M}{\partial {{r}_{+}}} \right)}_{P(\eta ),\beta ,\eta }}{{\left( \frac{\partial {{r}_{+}}}{\partial \eta } \right)}_{P(\eta ),\beta ,S}},
\end{align}
\begin{align}\label{3.13}
{{\left( \frac{\partial P}{\partial \eta } \right)}_{S,\beta ,M(\eta )}}&={{\left( \frac{\partial P}{\partial \eta } \right)}_{\beta ,{{r}_{+}},M(\eta )}}+{{\left( \frac{\partial P}{\partial {{r}_{+}}} \right)}_{\eta ,\beta ,M(\eta )}}{{\left( \frac{\partial {{r}_{+}}}{\partial \eta } \right)}_{S}}_{,\beta ,M(\eta )}.
\end{align}
When $M\to {{M}_{0}}$, $P\to {{P}_{0}}$, from Eqs. (\ref{3.12}) and (\ref{3.13}), we obtain
\begin{align}\label{3.14}
{{\left( \frac{\partial {{r}_{+}}}{\partial \eta } \right)}_{S}}_{,\beta ,M(\eta )}&=-\frac{\frac{\partial M}{\partial \eta }}{\frac{\partial M}{\partial {{r}_{+}}}},~~{{\left( \frac{\partial {{r}_{+}}}{\partial \eta } \right)}_{P(\eta ),\beta ,S}}=-\frac{\frac{\partial P}{\partial \eta }}{\frac{\partial P}{\partial {{r}_{+}}}}.
\end{align}
Substituting Eqs. (\ref{3.14}) into Eqs. (\ref{3.12}) and (\ref{2.14}), respectively, yields
\begin{align}\label{3.15}
{{\left( \frac{\partial P}{\partial \eta } \right)}_{S,\beta ,M(\eta )}}={{\left( \frac{\partial P}{\partial \eta } \right)}_{\beta ,{{r}_{+}},M(\eta )}}-{{\left( \frac{\partial P}{\partial {{r}_{+}}} \right)}_{\eta ,\beta ,M(\eta )}}\frac{\frac{\partial M}{\partial \eta }}{\frac{\partial M}{\partial {{r}_{+}}}}&=\frac{{{\left( \frac{\partial P}{\partial \eta } \right)}_{\beta ,{{r}_{+}},M(\eta )}}\frac{\partial M}{\partial {{r}_{+}}}-{{\left( \frac{\partial P}{\partial {{r}_{+}}} \right)}_{\eta ,\beta ,M(\eta )}}\frac{\partial M}{\partial \eta }}{\frac{\partial M}{\partial {{r}_{+}}}},
\end{align}
\begin{align}\label{3.16}
{{\left( \frac{\partial M}{\partial \eta } \right)}_{P(\eta ),\beta ,S}}={{\left( \frac{\partial M}{\partial \eta } \right)}_{P(\eta ),\beta ,{{r}_{+}}}}-{{\left( \frac{\partial M}{\partial {{r}_{+}}} \right)}_{P(\eta ),\beta ,\eta }}\frac{\frac{\partial P}{\partial \eta }}{\frac{\partial P}{\partial {{r}_{+}}}}&=\frac{{{\left( \frac{\partial M}{\partial \eta } \right)}_{P(\eta ),\beta ,{{r}_{+}}}}\frac{\partial P}{\partial {{r}_{+}}}-{{\left( \frac{\partial M}{\partial {{r}_{+}}} \right)}_{P(\eta ),\beta ,\eta }}\frac{\partial P}{\partial \eta }}{\frac{\partial P}{\partial {{r}_{+}}}}.
\end{align}

It has been demonstrated that the first law of thermodynamics is satisfied by the thermodynamic quantities of the black hole, we obtain
\begin{align}\label{3.17}
\delta M&=T\delta S+{{\psi }_{\Lambda }}\delta P+{{\psi }_{\beta }}\delta B,
\end{align}
with $B=\frac{\beta }{16\pi }$ \cite{S01078,S6251,D235014},
\begin{align}\label{3.18}
{{\psi }_{\beta }}&=-\frac{64\pi }{r_{+}^{5}}-\frac{832\pi {{(1+\eta )}^{4}}{{\Lambda }^{4}}r_{+}^{3}}{81}+\frac{2560\pi {{(1+\eta )}^{3}}{{\Lambda }^{3}}{{r}_{+}}}{27}-\frac{128\pi {{(1+\eta )}^{2}}{{\Lambda }^{2}}}{{{r}_{+}}}+\frac{256\pi (1+\eta )\Lambda }{3r_{+}^{3}} \notag \\
&+\beta \left( \frac{143360\pi {{(1+\eta )}^{3}}{{\Lambda }^{3}}}{9r_{+}^{5}} \right.+\frac{3072\pi }{r_{+}^{11}}-\frac{23552\pi {{(1+\eta )}^{2}}{{\Lambda }^{2}}}{r_{+}^{7}}+\frac{1638\pi (1+\eta )\Lambda }{r_{+}^{9}}\notag \\
&-\frac{99328\pi {{(1+\eta )}^{4}}{{\Lambda }^{4}}}{27r_{+}^{3}}+\frac{13312\pi {{(1+\eta )}^{6}}{{\Lambda }^{6}}{{r}_{+}}}{27}\left. -\frac{4096\pi {{(1+\eta )}^{5}}{{\Lambda }^{5}}}{9{{r}_{+}}} \right)+o({{\beta }^{2}}).
\end{align}
From Eqs. (\ref{3.15}), (\ref{3.16}) and (\ref{3.17}), we obtain
\begin{align}\label{3.19}
{{\left( \frac{\partial {{M}_{0}}}{\partial \eta } \right)}_{P(\eta ),\beta ,S}}&=-\underset{M\to {{M}_{0}}}{\mathop{\lim }}\,\frac{\frac{\partial M}{\partial {{r}_{+}}}}{\frac{\partial P}{\partial {{r}_{+}}}}{{\left( \frac{\partial P}{\partial \eta } \right)}_{S,\beta ,M(\eta )}}=-\underset{M\to {{M}_{0}}}{\mathop{\lim }}\,{{\psi }_{\Lambda }}{{\left( \frac{\partial P}{\partial \eta } \right)}_{S,\beta ,M(\eta )}}.
\end{align}

In the case of constant $\beta$ and $P$, $M$ is a function of $r_+$ and $\eta$, and from Eq. (\ref{2.13}), $S$ is also a function of $r_+$ and $\eta$. According to the law of derivation of composite function, we obtain
\begin{align}\label{3.20}
{{\left( \frac{\partial M}{\partial \eta } \right)}_{P,\beta ,S(\eta )}}&={{\left( \frac{\partial M}{\partial \eta } \right)}_{P,\beta ,{{r}_{+}}}}+{{\left( \frac{\partial M}{\partial {{r}_{+}}} \right)}_{P,\beta ,\eta }}{{\left( \frac{\partial {{r}_{+}}}{\partial \eta } \right)}_{P,\beta ,S(\eta )}},
\end{align}
\begin{align}\label{3.21}
{{\left( \frac{\partial S}{\partial \eta } \right)}_{P,\beta ,M(\eta )}}&={{\left( \frac{\partial S}{\partial \eta } \right)}_{P,\beta ,{{r}_{+}}}}+{{\left( \frac{\partial S}{\partial {{r}_{+}}} \right)}_{P,\beta ,\eta }}{{\left( \frac{\partial {{r}_{+}}}{\partial \eta } \right)}_{P,\beta ,M(\eta )}}.
\end{align}
When $M\to {{M}_{0}}$, $S\to {{S}_{0}}$, from Eqs. (\ref{3.12}) and (\ref{3.13}), we obtain
\begin{align}\label{3.22}
{{\left( \frac{\partial {{r}_{+}}}{\partial \eta } \right)}_{S}}_{,\beta ,M(\eta )}&=-\frac{\frac{\partial M}{\partial \eta }}{\frac{\partial M}{\partial {{r}_{+}}}},~~{{\left( \frac{\partial {{r}_{+}}}{\partial \eta } \right)}_{P,\beta ,S(\eta )}}=-\frac{\frac{\partial S}{\partial \eta }}{\frac{\partial S}{\partial {{r}_{+}}}}.
\end{align}
Substituting Eqs. (\ref{3.22}) into Eqs. (\ref{3.20}) and (\ref{3.21}), respectively, yields
\begin{align}\label{3.23}
{{\left( \frac{\partial {{M}_{0}}}{\partial \eta } \right)}_{P,\beta ,S(\eta )}}&=-\underset{M\to {{M}_{0}}}{\mathop{\lim }}\,\frac{\frac{\partial M}{\partial {{r}_{+}}}}{\frac{\partial S}{\partial {{r}_{+}}}}{{\left( \frac{\partial S}{\partial \eta } \right)}_{S(\eta ),\beta ,M(\eta )}}=-\underset{M\to {{M}_{0}}}{\mathop{\lim }}\,T{{\left( \frac{\partial S}{\partial \eta } \right)}_{S(\eta ),\beta ,M(\eta )}}.
\end{align}
The GP relation between the black hole energy $M$ and the thermodynamic state parameters $P$ and $S$ is given by Eqs (\ref{3.19}) and (\ref{3.23}), respectively. It is evident from the aforementioned calculations that the application of the derivation law of composite function is a more efficient approach compared to the direct calculation method. However, when calculating the GP relation, one of the thermodynamic state parameters of the black hole is taken as a perturbation parameter $\eta$. For example, $S$ or $P$ is taken as a function of the perturbation parameter $\eta$ when calculating the GP relation above, while the other parameters are discussed as invariants. In practical problems, each thermodynamic state parameter is a function of the perturbation parameter $\eta$, and it is necessary to find the universal GP relation when multiple thermodynamic state parameters are functions of the perturbation parameter $\eta$.

$3)$ full differential method

In the context of employing the composite function derivation method, it is imperative to note that the GP relation between each thermodynamic state parameter and the energy can only be calculated separately when the energy is a function of multiple thermodynamic state parameters. In the following, we employ the full differential method to compute the GP relation when multiple thermodynamic state parameters vary with the perturbation parameter, and the universal GP relation when multiple thermodynamic state parameters vary with the perturbation parameter $\eta$ is given.
As demonstrated in Eqs. (\ref{2.12}), (\ref{2.13}) and (\ref{2.14}), the energy of the black hole, $M$, is a function of $S(\eta )$, $P(\eta )$, $B(\eta )$ and $\eta $. Therefore, we obtain
\begin{align}\label{3.24}
{{\left( \frac{dM}{d\eta } \right)}_{S(\eta ),P(\eta ),B(\eta )}}&={{\left( \frac{\partial M}{\partial \eta } \right)}_{P(\eta ),B(\eta ),S(\eta )}}+{{\left( \frac{\partial M}{\partial S} \right)}_{P(\eta ),B(\eta ),\eta }}{{\left( \frac{\partial S}{\partial \eta } \right)}_{P(\eta ),B(\eta ),M(\eta )}} \notag \\
&+{{\left( \frac{\partial M}{\partial P} \right)}_{S(\eta ),B(\eta ),\eta }}{{\left( \frac{\partial P}{\partial \eta } \right)}_{S(\eta ),B(\eta ),M(\eta )}}+{{\left( \frac{\partial M}{\partial B} \right)}_{S(\eta ),B(\eta ),\eta }}{{\left( \frac{\partial B}{\partial \eta } \right)}_{P(\eta ),S(\eta ),M(\eta )}},
\end{align}
when the energy of the black hole $M\to {{M}_{0}}$, from Eqs. (\ref{3.17}) and (\ref{3.24}), we obtain
\begin{align}\label{3.25}
0&={{\left( \frac{\partial M}{\partial \eta } \right)}_{P(\eta ),B(\eta ),S(\eta )}}+T{{\left( \frac{\partial S}{\partial \eta } \right)}_{P(\eta ),B(\eta ),M(\eta )}}+{{\psi }_{\Lambda }}{{\left( \frac{\partial P}{\partial \eta } \right)}_{S(\eta ),B(\eta ),M(\eta )}}+{{\psi }_{\beta }}{{\left( \frac{\partial B}{\partial \eta } \right)}_{P(\eta ),S(\eta ),M(\eta )}},
\end{align}
namely,
\begin{align}\label{3.26}
{{\left( \frac{\partial {{M}_{0}}}{\partial \eta } \right)}_{P(\eta ),B(\eta ),S(\eta )}}&=\underset{M\to {{M}_{0}}}{\mathop{\lim }}\,-T{{\left( \frac{\partial S(\eta )}{\partial \eta } \right)}_{P(\eta ),B(\eta ),M(\eta )}} \notag \\
&-\underset{M\to {{M}_{0}}}{\mathop{\lim }}\,\left[ {{\psi }_{\Lambda }}{{\left( \frac{\partial P(\eta )}{\partial \eta } \right)}_{S(\eta ),B(\eta ),M(\eta )}}+{{\psi }_{\beta }}{{\left( \frac{\partial B(\eta )}{\partial \eta } \right)}_{P(\eta ),S(\eta ),M(\eta )}} \right].
\end{align}
When $P(\eta )$ and $B(\eta )$ are invariant, Eq. (\ref{3.26}) reduces to
\begin{align}\label{3.27}
{{\left( \frac{\partial {{M}_{0}}}{\partial \eta } \right)}_{P,B,S(\eta )}}&=-\underset{M\to {{M}_{0}}}{\mathop{\lim }}\,T{{\left( \frac{\partial S(\eta )}{\partial \eta } \right)}_{P,B,M(\eta )}},
\end{align}
when $S(\eta )$ and $B(\eta )$ are invariant, Eq. (\ref{3.26}) reduces to
\begin{align}\label{3.28}
{{\left( \frac{\partial {{M}_{0}}}{\partial \eta } \right)}_{P(\eta ),B,S}}&=-\underset{M\to {{M}_{0}}}{\mathop{\lim }}\,{{\psi }_{\Lambda }}{{\left( \frac{\partial P(\eta )}{\partial \eta } \right)}_{S,B,M(\eta )}},
\end{align}
when $S(\eta )$ and $P(\eta )$ are invariant, Eq. (\ref{3.26}) reduces to
\begin{align}\label{3.29}
{{\left( \frac{\partial {{M}_{0}}}{\partial \eta } \right)}_{P,B(\eta ),S}}=-\underset{M\to {{M}_{0}}}{\mathop{\lim }}\,{{\psi }_{\beta }}{{\left( \frac{\partial B(\eta )}{\partial \eta } \right)}_{P,S,M(\eta )}},
\end{align}
when $P(\eta )$ is invariant, Eq. (\ref{3.26}) reduces to
\begin{align}\label{3.30}
{{\left( \frac{\partial {{M}_{0}}}{\partial \eta } \right)}_{P,B(\eta ),S(\eta )}}&=-\underset{M\to {{M}_{0}}}{\mathop{\lim }}\,\left[ T{{\left( \frac{\partial S(\eta )}{\partial \eta } \right)}_{P,B(\eta ),M(\eta )}}+{{\psi }_{\beta }}{{\left( \frac{\partial B(\eta )}{\partial \eta } \right)}_{P,S(\eta ),M(\eta )}} \right],
\end{align}
when $B(\eta )$ is invariant, Eq. (\ref{3.26}) reduces to
\begin{align}\label{3.31}
{{\left( \frac{\partial {{M}_{0}}}{\partial \eta } \right)}_{P(\eta ),B,S(\eta )}}&=-\underset{M\to {{M}_{0}}}{\mathop{\lim }}\,\left[ T{{\left( \frac{\partial S(\eta )}{\partial \eta } \right)}_{P(\eta ),B,M(\eta )}}+{{\psi }_{\Lambda }}{{\left( \frac{\partial P(\eta )}{\partial \eta } \right)}_{S(\eta ),B,M(\eta )}} \right],
\end{align}
when $S(\eta )$ is invariant, Eq. (\ref{3.26}) reduces to
\begin{align}\label{3.32}
{{\left( \frac{\partial {{M}_{0}}}{\partial \eta } \right)}_{P(\eta ),B(\eta ),S}}&=-\underset{M\to {{M}_{0}}}{\mathop{\lim }}\,\left[ {{\psi }_{\Lambda }}{{\left( \frac{\partial P(\eta )}{\partial \eta } \right)}_{S,B(\eta ),M(\eta )}}+{{\psi }_{\beta }}{{\left( \frac{\partial B(\eta )}{\partial \eta } \right)}_{P(\eta ),S,M(\eta )}} \right].
\end{align}

According to Eq. (\ref{3.27}), in the event that the state parameters $P$ and $B$ of the black hole are not functions of the perturbation parameter $\eta$, the expressions derived are the currently accepted expressions for the GP relation. However, it is evident from the Eqs. (\ref{2.12}) and (\ref{2.13}) that Eq. (\ref{3.27}) for the black hole energy $M$ and entropy $S$ contain higher-order terms for the perturbation parameter $\eta$. The generalized GP relation, as presented by Eq. (\ref{3.26}), is obtained, which not only contains the relation of energy $M$ and entropy $S$ with the variation of perturbation parameter $\eta$, but also the universal GP relation contains $P$ and $B$ with the variation of perturbation parameter $\eta$.

\section{Summary}\label{four}

The present study is grounded on the following core idea: Given the thermodynamic quantities of black holes, in accordance with the constraints imposed by Eq. (\ref{2.12}), the thermodynamic quantity and the horizon position of black hole will concomitantly be modified when the perturbation parameter of the spacetime, $\eta$, is altered. This physical mechanism not only verifies the salient conclusion of the Ref. \cite{G101103}, but also generalizes it to the following two more general types of scenarios: Firstly, the complex spacetime situation in which the thermodynamic quantities of the black hole are a higher-order function of the perturbation parameter $\eta$. Secondly, the complex spacetime situation in which the energy, $M$, of the black hole is a function of a multi-parameter ($S$, $P$, $B$, etc.). A notable strength of the present derivation is its computational simplicity and logical clarity, which renders it a reliable theoretical instrument for subsequent in-depth studies.

A thorough examination of the proportional relationship between modified masses and thermodynamic quantities has the potential to reveal novel avenues for understanding the Weak Gravity Conjecture (WGC). As is well known, the WGC offers significant physical insights into the theory of quantum gravity. The findings of this study are expected to further refine our comprehension of this foundational physical problem.

In summary, the primary innovations of this investigation are twofold: (1) revealing the universal relation when the energy $M$ as a function of multiple thermodynamic quantities, and (2) formulating the generalized GP relation when each thermodynamic quantity is a function of a higher-order perturbation parameter, $\eta$. These theoretical findings not only provide new perspectives for comprehending the WGC but also contribute to the profound advancement of quantum gravity theory. It is anticipated that these findings will provide significant theoretical foundations and research directions for future studies in this field.

\section*{Acknowledgments}
We would like to thank Prof. Ren Zhao and Meng-Sen Ma for their indispensable discussions and comments. This work was supported by the Natural Science Foundation of China (Grant No. 12375050, Grant No. 11705106, Grant No. 12075143), the Scientific Innovation Foundation of the Higher Education Institutions of Shanxi Province (Grant Nos. 2023L269)

\end{document}